\newcommand{\du}{\,cm$^{-3}$\,}

\newcommand{\fu}{\,erg cm$^{-2}$ s$^{-1}$\,}

\documentclass[apjl]{emulateapj}
\usepackage{natbib}
\usepackage{apjfonts}

\shortauthors{Maggio \& Ness}
\shorttitle{Density variations in the corona of AD Leo}

\begin{document}

\title{Spectral indications of density variability in the corona of AD Leonis}

\author{A. Maggio\altaffilmark{1}, J.-U. Ness\altaffilmark{2}}

\altaffiltext{1}{INAF -- Osservatorio Astronomico di Palermo,
Piazza del Parlamento 1, I--90134 Palermo, Italy;
maggio@astropa.unipa.it.}
\altaffiltext{2}{Department of Physics, Rudolf Peierls Centre for
Theoretical Physics, University of Oxford, 1 Keble
Road, Oxford OX1\,3NP; ness@thphys.ox.ac.uk}

\begin{abstract}

Direct comparison of high-resolution X-ray spectra of the
active dMe star AD Leo, observed three times with
{\em Chandra}, shows variability of key density-sensitive lines,
possibly due to flaring activity. In particular, a significant
long-duration enhancement of the coronal density is indicated by
the \ion{Ne}{9} and \ion{Si}{13} He-like triplets, and possibly also by
density-sensitive \ion{Fe}{21} line ratios.

\end{abstract}

\keywords{stars: individual (AD Leo) --- stars: activity --- stars: coronae
--- stars: late-type --- X-rays: stars}

\clearpage

\section{Introduction}
\label{sec:intro}

The plasma density is a crucial parameter for understanding the structure of
stellar coronae. Rather than being uniformly distributed on large 
spatial scales, as occurs in the photosphere, the coronal plasma
is trapped by surface magnetic fields within scale lengths typically
much smaller than the stellar radius, and is likely 
to be very inhomogeneous in density. The patchy appearance of
the solar corona in X-rays is mainly a density effect; since the radiative
emission of an optically thin plasma excited by collisional excitation is
proportional to the square density, the observed irradiance traces
preferentially high-density ``compact'' regions, confined by small-scale
magnetic fields. The plasma density is also expected to vary
in time owing to the dynamics of these fields and of the plasma itself; 
in particular, the coronal density is expected to increase {\it locally}
as a consequence of impulsive heating events (flares) which trigger an upward
motion of denser plasma from deeper atmospheric layers (``chromospheric
evaporation'').

The measurement of plasma densities in stellar coronae is a relatively
recent practice in X-ray spectroscopy; Chandra and XMM-Newton 
high-resolution gratings provide us with a number of density-sensitive
emission lines, which are now routinely employed to this purpose.
However, density estimates and their interpretation are more
difficult and challenging than expected, for several reasons; first,
even in relatively strong X-ray sources the relevant emission lines may
be difficult to measure owing to low ion abundances, severe blends with
other spectral features, or insufficient instrumental sensitivity in the
relevant wavelength ranges; secondly, different spectral diagnostics
probe coronal regions at different temperatures, and they often yield
density estimates which are difficult to reconcile; finally, the lack
of spatial resolution in observations
of stellar coronae allows us to obtain only ``effective'' densities, averaged
over all the coronal structures in the visible hemisphere and weighted by
the plasma irradiance, so that any interpretation becomes model dependent.

For the above reasons, it is currently easier to look for 
variations with time in the plasma density in a given coronal source, 
rather than comparing absolute values derived from snapshots of
different stars. In this respect, energetic events like stellar flares
are natural experiments to examine, and they have
been the subject of several investigations in recent years; nonetheless,
convincing evidence of density variations are scanty in the literature,
with notable exceptions such as the flare recently observed in the nearby
dMe star Proxima Centauri \citep{gash02}.

In this paper we present the first evidence of significant long-term 
density variations in the corona of AD Leo, another well-known dMe flare star 
observed several times with the {\it Chandra} X-ray Observatory
\citep{wbc+02}. The strength of the present evidence comes from the
direct comparison of high-resolution X-ray spectra taken at different
times, with little need of data adjustments or problematic atomic models.
On the other hand, the interpretation of the results is unclear,
but gives some new hints on issues under debate in stellar X-ray
spectroscopy and coronal physics.

\section{Target and Observations}
\label{sec:data}

AD~Leo is one of the X-ray brightest single dMe stars currently known,
having quite a stable quiescent X-ray luminosity, 
$L_{\rm x} = 3$--$5 \times 10^{28}$\,erg s$^{-1}$ (0.5--4.5\,keV band)
over a 17\,yr period, as determined by \citet{fmr00}. On the other hand,
it is also a quite variable X-ray source on short time scales, owing to
its characteristic coronal magnetic activity \citep[see e.g.][]{agdk00}.

AD~Leo was observed twice with the {\it Chandra} Low-Energy Transmission
Grating Spectrometer (LETGS) 
in January and October 2000, as part of the GTO program, and once again 
in June 2002 with the High-Energy Transmission
Grating Spectrometers (HETGS). A detailed analysis of the October 2000
observation was presented by \citet{mdk+04}, including a reconstruction
of the plasma emission measure distribution (EMD) with temperature, and
evaluation of the plasma densities at various temperatures by means of the
He-like triplets and \ion{Fe}{21} lines. Here we will focus on the
comparison of this observation with the previous one, since they have
about the same
exposure time (48\,ks and 46\,ks, respectively), and we will also
briefly discuss results based on the shorter Jan 2000 observation (10\,ks).

The data were retrieved from the {\it Chandra} archive and re-processed 
with the {\it Chandra} Interactive Analysis of Observations (CIAO V3.1) 
software.
We used the tool {\sc fullgarf} to calculate effective detector areas.
From each observation we have extracted either the LETGS spectrum 
(6--140\,\AA), or both the first-order
Medium-Energy Grating (MEG, 1--25\,\AA) and High-Energy Grating (HEG,
1--18\,\AA) spectra, simultaneously provided by the HETGS; photons from the two
dispersion directions were coadded to maximize the S/N ratio.
We also extracted the spectra in consecutive time bins in order to
create the light curves, as described in the following section.

\begin{figure}[!ht]
   \resizebox{\hsize}{!}{
   \plotone{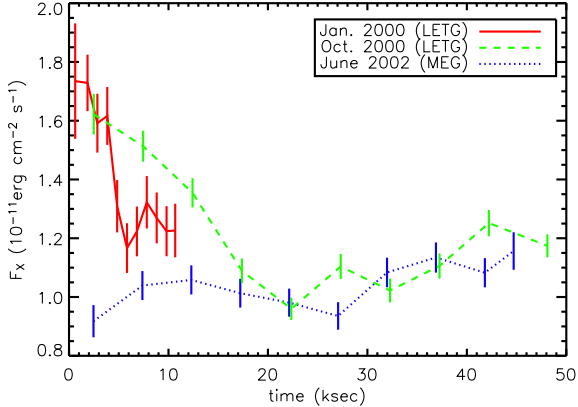}
                        }
   \caption{
AD Leo X-ray light curves (net observed flux in the 7--25\,\AA\ wavelength
range) derived from the {\it Chandra} observations reported in the
legend. The fluxes were computed by integrating the dispersed spectra
extracted in equal time bins, within each observation. 
   \label{fig:lc}}
\end{figure}

\section{Analysis and Results}
\label{sec:a+r}

\subsection{X-ray light curve}
\label{sec:lc}

Figure~\ref{fig:lc} shows the X-ray light curves obtained from the three 
{\it Chandra} observations: dispersed spectra were extracted from
successive time intervals of equal duration, and total X-ray
fluxes were computed in the common wavelength range 7--25\,\AA, after
background subtraction and correction for the instrument effective area 

No strong isolated flare is visible during these observations, but the
Jan and Oct 2000 light curves both show an initial emission level
higher than the quiescent one ($f_{\rm x} \approx 10^{-11}$\fu) by a
factor $\la 2$, followed by a clear decline; this behavior resembles
the decay phase of a flare, but we note that in the Oct 2000 light curve 
the decline is much slower than the expected exponential rate.
The latter observation was recently studied by
\citet{brm+03}, who performed a separate analysis of the first 12\,ks
segment and concluded that the high-temperature tail of EMD was enhanced with
respect to the following time segment; on the other hand, \citet{mdk+04}
argue that a rotational modulation effect cannot be completely excluded because 
the duration of the October 2000 observation is a 
sizable fraction ($\sim 20$\%) of the photometric rotational period 
\citep[$P_{\rm rot} = 2.7$\,d,][]{sh86}.

\begin{figure*}[!ht]
   \centering
   \resizebox{0.90\hsize}{!}{
   \plottwo{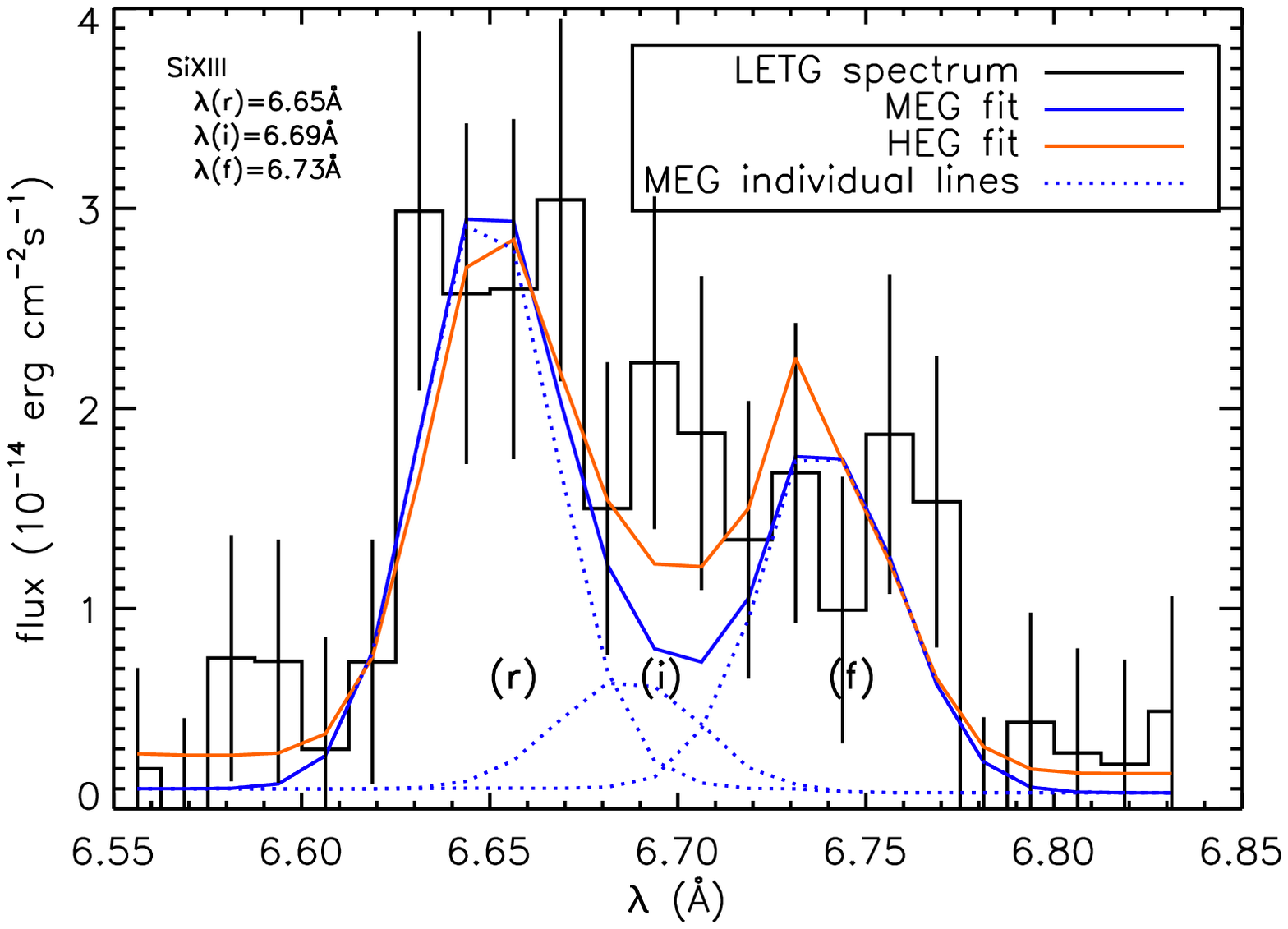}{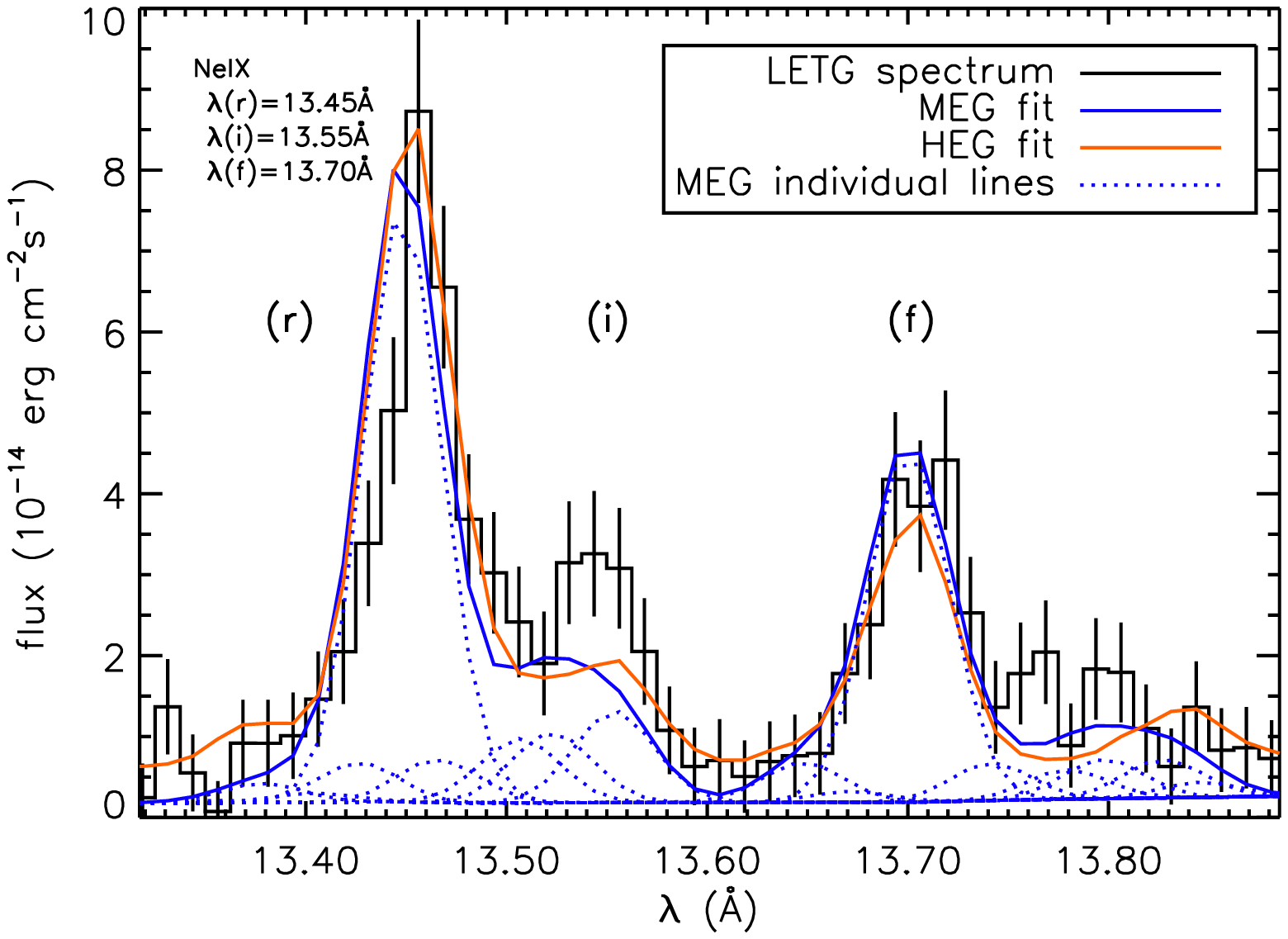}
                        }
   \resizebox{0.90\hsize}{!}{
   \plottwo{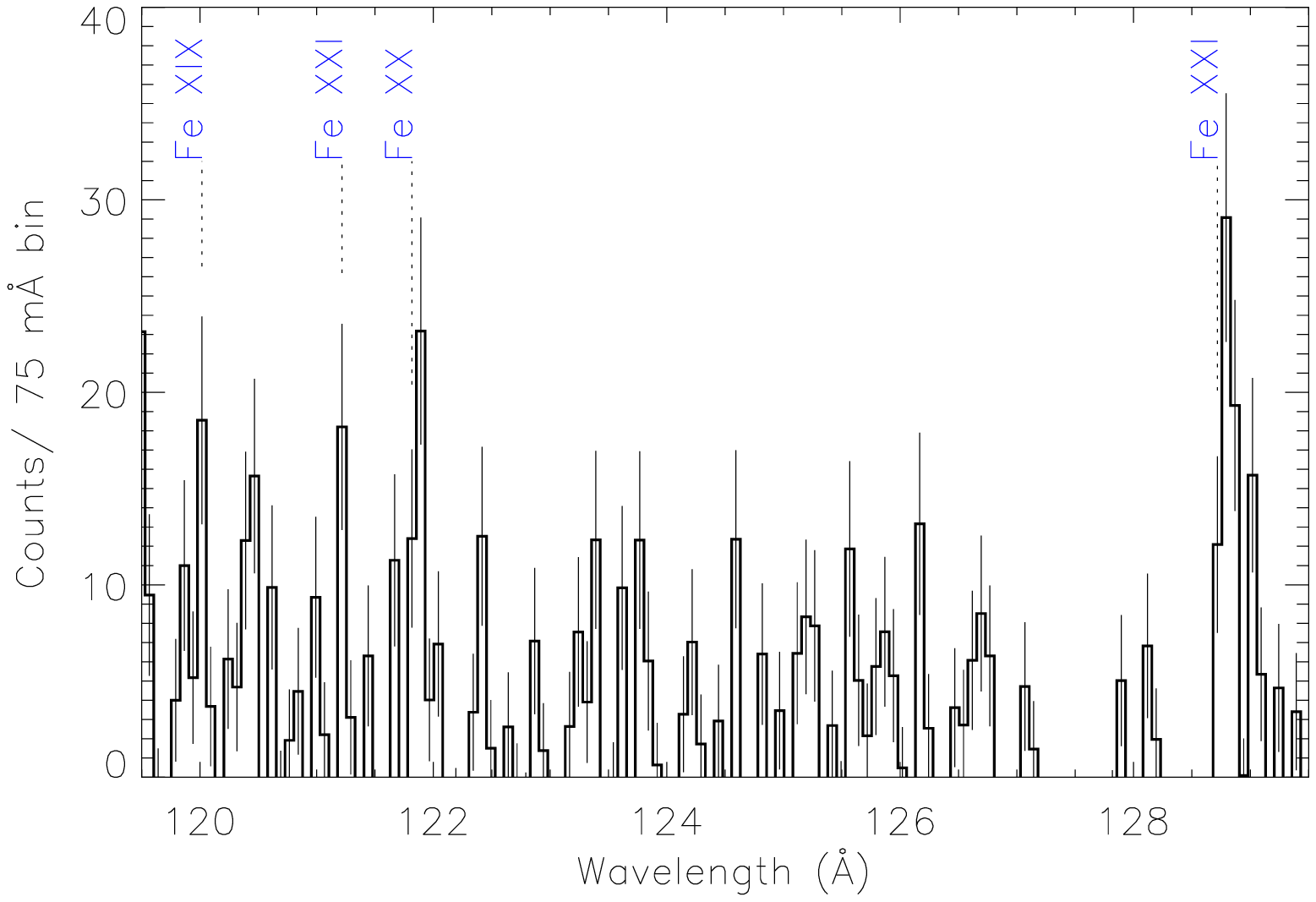}{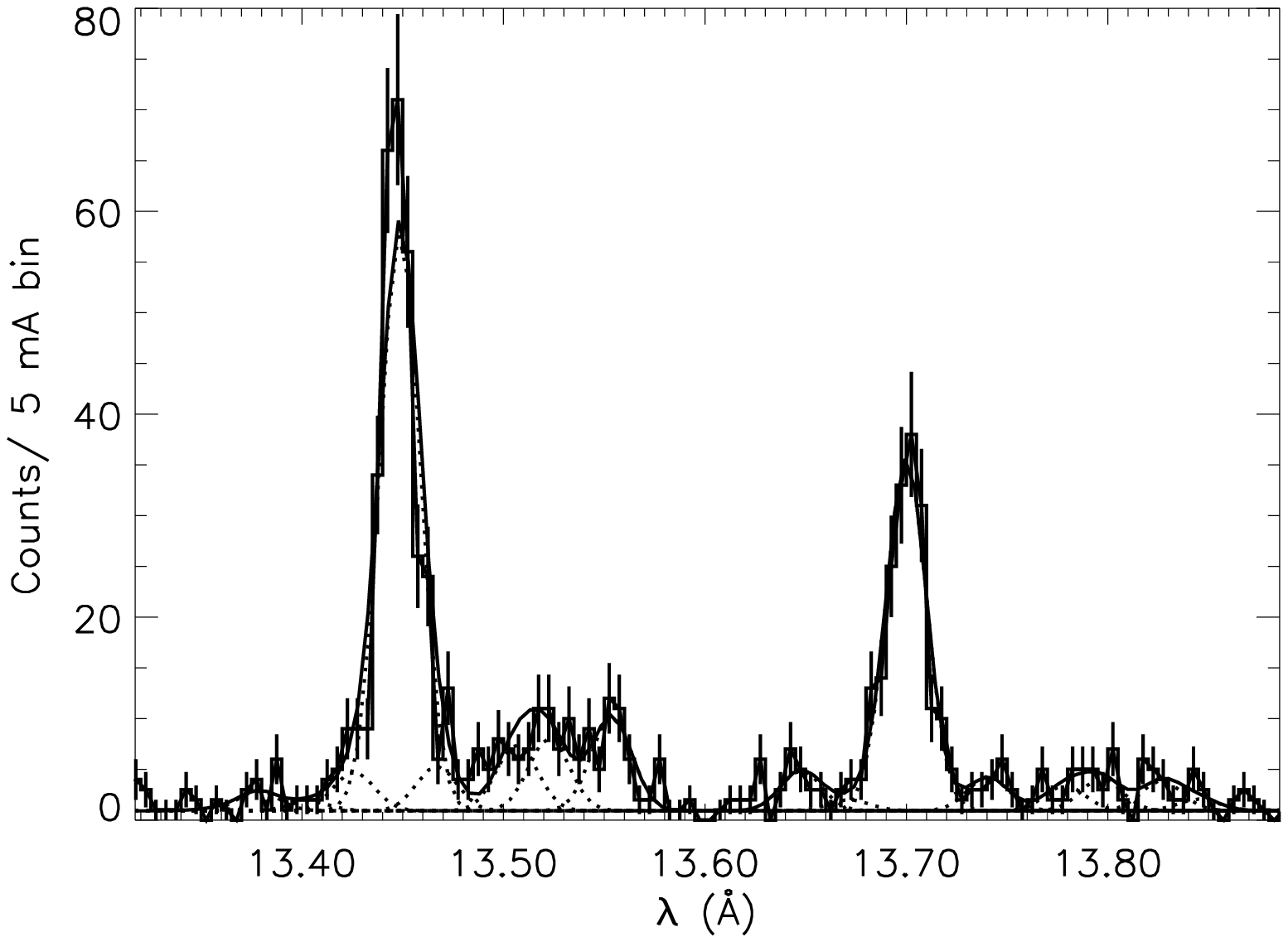}
   }
\caption{
Upper panels: LETG \ion{Si}{13} and \ion{Ne}{9} triplets compared with
scaled HETGS line fitting models. Lower left: LETG spectral region
including \ion{Fe}{21} lines at 121.2\AA\ and 128.7\AA. 
Lower right: \ion{Ne}{9} triplet in the MEG spectrum of AD\,Leo,
with 18 best-fit line components.
\label{fig:sp}}
\end{figure*}

\subsection{Analysis and Comparison of Spectra}
\label{sec:sp}

We have focused our attention on the spectral ranges including the
\ion{Si}{13}, \ion{Mg}{11}, \ion{Ne}{9}, and \ion{O}{7} He-like
triplets, and on selected density-sensitive \ion{Fe}{21} lines.
Note that the first three triplets are visible in all LETGS and HETGS
spectra, while the \ion{O}{7} triplet can be observed only with the
LETGS and the HETGS/MEG. With regard to the \ion{Fe}{21} lines, we have
considered both $2s^22p3d$ transitions to the $2s^22p^2$ ground state
which fall at wavelengths in the range 12.28--12.53\,\AA, and transitions
from $2s2p^3$ states which are visible at wavelengths 102.2--145.7\,\AA\
only in LETGS spectra \citep[for more details]{ngs+04}.

For the comparison of LETGS and HETGS spectra (Fig.~\ref{fig:sp})
we proceeded as follows. First, we have fitted the relevant lines
in the MEG and HEG spectra using the program CORA \citep{nw02};
the smooth representation of each spectral region obtained with the
above line fitting was converted to energy fluxes by use of the effective
areas and exposure times, and finally rebinned to the resolution of the
LETGS spectra; this procedure allows the proper overlay of the HETGS and LETGS 
data for ease of direct comparison.

Note that the line fitting is performed with a maximum likelihood method
applied to the total source+background spectrum. In each spectral region
the strongest lines are fitted with normalized Moffat profiles (Lorentzians
with an exponent $\beta=2.4$), representing the instrumental line spread
function. The
instrumental background is not subtracted from the total measured spectrum,
but is instead added to the model spectrum consisting of the sum of all
line emission components. The line counts are obtained as normalization factors
of the line profiles.

Figure \ref{fig:sp} shows the HETGS vs. LETGS (October 2000) comparison
for the \ion{Si}{13} and \ion{Ne}{9} triplets; the original
HETGS/MEG spectrum in the region of \ion{Ne}{9} is also shown.
What is immediately evident is the excess emission at the positions of
the \ion{Si}{13} and \ion{Ne}{9} intercombination lines 
(6.69\,\AA\ and 13.55\,\AA, respectively) in the LETGS
spectrum with respect to the HETGS case. No such enhancement is
visible in the \ion{O}{7} (21.6, 21.8, 22.1\,\AA) and \ion{Mg}{11}
(9.17, 9.23, 9.31\,\AA) triplets (not shown); 
note that the latter is affected by low signal to noise ratio because of the
relatively low Mg abundance in the corona of AD~Leo \citep{mdk+04}.

\begin{deluxetable}{rrrrrrr}
\tablewidth{0pt}
\tablecaption{Density-sensitive \ion{Fe}{21} lines
\label{tab:fe21}}
\tablecolumns{7}
\tablehead{
\multicolumn{1}{c}{$\lambda$} & \multicolumn{1}{c}{$\log N_{\rm e}^{\rm c}$} &
       & \multicolumn{1}{c}{$A_{\rm eff}$} & ISM &
$\lambda/128.73$ & \multicolumn{1}{c}{$\log N_{\rm e}$} \\
\multicolumn{1}{c}{[\AA]}     & \multicolumn{1}{c}{[cm$^{-3}$]} & 
Counts & \multicolumn{1}{c}{[cm$^2$]}      & ${\rm e}^{-\tau}$ &
line ratio       & \multicolumn{1}{c}{[cm$^{-3}$]}
}
\startdata
 97.88 & 12.0~$\uparrow$ & $< 10$     & 7.08 & 0.91 & $< 0.13$ & $< 12.3$ \\
102.22 & 12.0~$\uparrow$ & $18 \pm 8$ & 6.71 & 0.90 & $0.19 \pm 0.10$ & $< 12.3$ \\
117.51 & 13.5~$\downarrow$ & $<  8$     & 6.17 & 0.85 & $< 0.13$ & $< 11$ \\
121.21 & 12.4~$\uparrow$ & $22 \pm 9$ & 5.57 & 0.84 & $0.30 \pm 0.15$ & 12.0--12.9 \\
128.73 & 12.7~$\downarrow$ & $46 \pm 11$& 3.67 & 0.81 \\
142.16 & 12.7~$\uparrow$ &  $< 15$     & 3.87 & 0.76 & $< 0.43$ & $< 13.0$ \\
145.65 & 12.5~$\uparrow$ & $15 \pm 9$ & 3.69 & 0.75 & $0.35 \pm 0.23$ & 12.0--13.1 \\
\enddata
\tablecomments{Col 1: theoretical wavelength. Col 2: critical density,
and arrow indicating the emissivity trend for increasing $N_{\rm e}$.
Col 3: observed counts or upper limits. Col 4: LETGS effective area. 
Col 5: ISM absorption factor, assuming $N_{\rm H} = 3\times10^{18}$\,cm$^{-2}$ 
\citep{cfh+97}. Col 6: photon flux ratio. Col 7: estimated electron density.}
\end{deluxetable}

We have verified that the above enhancements are still visible, although
somewhat reduced, also in the last 32\,ks of the October 2000
observations, i.e. where the X-ray emission is again at its quiescent level.
Some enhancement is also visible in the \ion{Ne}{9} intercombination
line of the January 2000 LETGS spectrum, although the shorter exposure
time makes the signal to noise ratio too low for any robust indication.

\begin{figure}[!ht]
\plotone{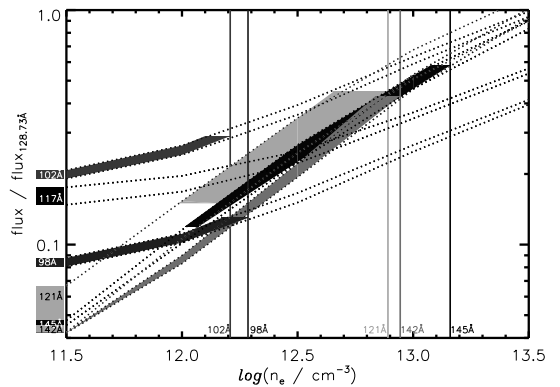}
\caption{
\ion{Fe}{21} predicted density-sensitive \ion{Fe}{21} line ratios,
for the temperature range $10^{6.8}$--$10^{7.2}$\,K (lower and upper
borders). The shaded areas indicate the density ranges consistent
with the data;
the vertical lines are placed at the high-density limits.
\label{fig:lratio}}
\end{figure}

We have also carefully re-examined the long-wavelength ($\lambda >
100$\,\AA) region of the LETGS spectrum, for evidence of density-sensitive
\ion{Fe}{21} lines; in particular, we have searched the lines
listed in Table~\ref{tab:fe21}, and measured total counts or
upper limits, as appropriate. We confirm the 
detection of the strong 128.7\,\AA\ line (Fig.\,\ref{fig:sp}), and also of
weaker spectral features at $\lambda$ 102.2, 121.2, and 145.65\,\AA, 
already tentatively identified by \citet{mdk+04}. Unfortunately, all the
latter three features are characterized by quite a low S/N ratio
(2--3$\sigma$), and there
are residual uncertainties in the wavelength calibration of the LETGS that
suggest caution in their identification; they fall at the 
nominal position of density-sensitive \ion{Fe}{21} lines, and we will treat
them as such in the following.
The densities (or upper limits) implied by these
diagnostics were estimated from the predicted line ratios
(Fig.~\ref{fig:lratio}), according to APED
\citep[Plasma Emission Database V1.3,][]{sbl+01}; all the observed ratios,
properly corrected for the instrument effective area and ISM absorption
factor, consistently indicate a density $N_{\rm e}$ in the range 
$10^{12.0}$--$10^{12.2}$\du, 
within statistical uncertainties, with the exception of the
117.51/128.73 ratio. Note that the non-detection of the 117.51\,\AA\ line is
puzzling, because this line is predicted to be relatively strong and
very weakly dependent on the electron density, hence we suspect that
its theoretical emissivity is overestimated.

It is more difficult to measure the \ion{Fe}{21} line 
fluxes at short wavelengths because the spectral region 12.2--12.6\,\AA\
is especially crowded with hundreds of relatively faint iron lines, 
mostly from \ion{Fe}{19}--\ion{}{22}. In particular, we have 
searched for the density-sensitive lines at 12.284\,\AA, and 
12.327\,\AA. 
The former has the highest emissivity 
at densities below $N_{\rm e} \approx 10^{12.7}$\du and it is 
clearly present in all X-ray spectra we have analyzed, while
the second line is expected to become enhanced only 
in the high-density regime, above $N_{\rm e} \approx 10^{12.4}$\du,
and it is not convincingly detected in any of the spectra. 

\section{Discussion and conclusions}
\label{sec:discuss}

The measurement of coronal plasma densities by means of X-ray spectral 
diagnostics is a difficult task (see \S 1). For
example, the \ion{Ne}{9} triplet is heavily blended with \ion{Fe}{19}
lines, and to a lesser extent with \ion{Fe}{20}--\ion{}{21} lines.
A detailed analysis of this spectral region in {\it Chandra} and
XMM-Newton spectra of Capella \citep{nbdh03} showed that APED is
sufficiently accurate and complete that all
significant observed lines in this region can be reasonably identified
in the {\it Chandra}/HEG spectra, but spectra at lower resolution may be
inadequate to derive reliable results.
In fact, independent analyses of the \ion{Ne}{9} triplet in the same
(October 2000) LETGS spectrum of AD~Leo, performed by \citet{nsb+02} and
by \citet{mdk+04}, resulted in quite different estimates of the plasma
density owing to the different treatment of blends by these authors. 

An apparent inconsistency emerges also by comparing the results in
\citet{mdk+04} and in the recent work by \citet{ngs+04}, who
included the {\it Chandra}/HETGS observation of AD~Leo in a large sample of
grating spectra of stellar coronae. While the former paper reports
densities $N_{\rm e} \sim 4\times10^{11}$\du at $T \sim 10^{6.6}$\,K (peak
emissivity temperature of the \ion{Ne}{9}) and $N_{\rm e} \sim
2\times10^{12}$\du at $T \sim 10^7$\,K (\ion{Fe}{21}), no evidence of
especially high densities appears in the survey by \citet{ngs+04}.

The present investigation resolves this issue; the direct comparison of
the LETGS and HETGS spectra of AD~Leo suggests that the star was
observed in different coronal emission regimes, a high-density state in 
October 2000 and a low-density state in June 2002. The high density
is strongly supported by the enhanced flux observed at the positions
of the intercombination lines of the \ion{Ne}{9} and
\ion{Si}{13} He-like triplets; our investigation of
density-sensitive \ion{Fe}{21} lines in the October 2000 LETGS spectrum
appears to be consistent with the above conclusion, although there
remain uncertainties in the identification of some of these lines.
Note that densities of a $few \times 10^{12}$\du in the corona of AD~Leo
have already been reported by \citet{sm02}, based on EUVE spectra.

The above picture is not entirely clear for two reasons;
first, a density increase should affect both the intercombination ($i$) line
and the forbidden ($f$) line in each He-like triplet, while we see only an
enhancement of the $i$ line; secondly, the temperature-sensitive ratios
$G(T) = (i+f)/r$, with $r$ denoting the resonance line, 
is significantly higher 
in the October 2000 LETGS spectrum than in the HETGS spectrum, with the
consequence that, apparently, the plasma temperature was lower in the
high-density state, a result which is difficult to understand.
Since the \ion{Ne}{9} $i$ line is strongly blended with \ion{Fe}{19},
which forms at a higher temperature ($T \sim 10^{6.9}$\,K), 
we have checked whether the
enhancement of the $i$ line could be explained by a contamination
effect; none of the strongest lines from other comparably hot ions appears
to be significantly different in the two spectra, and hence we have
no reason to suspect an increased \ion{Fe}{19} line
emission in the Oct 2000 corona of AD~Leo. 
Another possibility is that of a cooling plasma in which 
the population of the $1s2p~^3P$ levels (and hence the $i$ line) is
preferentially enhanced with respect to the $^3S$ level ($f$ line) 
in case of strong recombination effects; however, the expected
recombination time scales are much shorter than the observation length,
hence a continuous replenishment of plasma is required to
explain our finding. We conclude that the temperature indicated by the
\ion{Ne}{9} $G(T)$ ratio is not reliable, for reasons yet to be
understood.

Another puzzling result comes from the comparison of
the \ion{O}{7} triplet in the different observations: contrary to naive
expectation, the density-sensitive $f/i$ ratio turns out to be
significantly higher in October 2000 than in June 2002 ($3.2 \pm 0.6$
vs.\ $1.9 \pm 0.6$, respectively), indicating a {\it lower} density 
in the first case ($\log N_{\rm e} = 9.9 \pm 0.7$ and $10.5 \pm 0.3$\du,
respectively). Instead, the temperature-sensitive $G(T)$ ratios
are in perfect agreement, and they always indicate an ``effective''
formation temperature of $\approx 2\times10^6$\,K, i.e. coincident with the
\ion{O}{7} emissivity maximum. Again, this is not easy to explain, because
the plasma emission measure distribution vs. temperature in the corona
of AD~Leo peaks at $\approx 8$\,MK, and hence one expects an important
contribution to the \ion{O}{7} line emission coming from plasma at
temperatures higher than 2\,MK. 
This anomalous behavior of the 
$G(T)$ ratios was also noted in the survey of coronal densities
by \citet{tdpd04}.

In conclusion,
our detailed comparison of the different {\it Chandra} observations of
AD~Leo indicates that the plasma density in the corona of this active star 
may change substantially. The coincidence of the observed density
enhancement with a higher X-ray emission level is suggestive of a
dynamic phenomenon, possibly related to flaring activity. 
However, the observed variability of the X-ray emission can be due also 
to active regions in the stellar corona coming in and out of sight,  
due to rotational modulation, or due to the natural lifetime of
the magnetic fields which provide confinement and heating of the plasma.
For this reason, a significant variation of the temperature may not
occur.

In any case, the interpretation of widely used spectral diagnostics,
such as the He-like triplets or \ion{Fe}{21} density-sensitive lines is
certainly complicated by the likely existence of density inhomogeneities
in the coronal plasma, besides obvious uncertainties in the atomic
physics and in our ability to resolve all the relevant spectral
components with the available data. In fact, as already pointed out by
\citet{g04}, the observed line ratios may not describe ``local
densities'' but rather the steepness of the density distribution with 
temperature.
In this respect, predictions of the coronal X-ray spectra, taking
into account physically realistic distributions of the magnetic
structures, would be the next step in achieving progress in
stellar coronal physics and X-ray spectroscopy.

\acknowledgements{AM acknowledges partial support from
Ministero dell'Universit\'a e della Ricerca Scientifica.
JUN acknowledges support from PPARC under grant number PPA/G/S/2003/00091.
We also thank C.\ Jordan and P.\ Testa for helpful discussions and comments.
}


\end{document}